\documentclass[9pt,twocolumn,twoside]{pnas-new}
\usepackage{mathtools, hyperref, amsmath, amsfonts}
\usepackage{bm, xcolor, graphicx, subcaption, cancel}

\definecolor{RLtrainDSM}{HTML}{4daf4a}
\definecolor{RLtrainSSM}{HTML}{984ea3}
\definecolor{RLtrainLL}{HTML}{377eb8}
\definecolor{RLtrainGERM}{HTML}{e41a1c}
\definecolor{RLtrainGRU}{HTML}{ff7f00}
\definecolor{RLtrainSINGLE}{HTML}{e6ab02}
\definecolor{mygreen}{rgb}{0.0, 0.0, 0.0}
\definecolor{myred}{rgb}{0.0, 0.0, 0.0}
\definecolor{myblue}{rgb}{0.0, 0.0, 0.0}
\def\dir{.}

\def\replyOne{\color{myblue}}
\def\replyTwo{\color{myred}}
\def\replyThr{\color{mygreen}}

\usepackage{scalerel,stackengine}
\stackMath
\newcommand\reallywidehat[1]{
	\savestack{\tmpbox}{\stretchto{%
			\scaleto{%
				\scalerel*[\widthof{\ensuremath{#1}}]{\kern-.6pt\bigwedge\kern-.6pt}
				{\rule[-\textheight/2]{1ex}{\textheight}}%WIDTH-LIMITED BIG WEDGE
			}{\textheight}
		}{0.5ex}}
	\stackon[1pt]{#1}{\tmpbox}
}
\templatetype{pnasresearcharticle}
\title{Automating Turbulence Modeling \\ by Multi-Agent Reinforcement Learning}

\author[a]{Guido Novati}
\author[a]{Hugues Lascombes de Laroussilhe} 
\author[a,1]{Petros Koumoutsakos}

\affil[a]{Computational Science and Engineering Laboratory, Clausiusstrasse~33, ETH~Z\"{u}rich, CH-8092, Switzerland}
\leadauthor{Novati} 

 %  $|$ ... 

\begin{abstract}
Turbulent flow models are critical for applications such as aircraft design, weather forecasting and climate prediction. Existing models are largely based on physical insight and engineering intuition. More recently, machine learning has been contributing to this endeavor with  promising results. However all efforts have focused on supervised learning, which encounters difficulties to generalize beyond the training data.

Here, we introduce multi-agent reinforcement learning (MARL) as an automated discovery tool of turbulence models. We demonstrate the potential of this approach on Large Eddy Simulations of isotropic turbulence, using as reward the recovery of the statistical properties of Direct Numerical Simulations. The closure model is a control policy enacted by cooperating agents, which detect critical spatio-temporal patterns in the flow field to estimate the unresolved sub-grid scale physics. Results obtained with MARL algorithms based on experience replay compare favorably with established modeling approaches. Moreover, we show that the learned turbulence models generalize across grid sizes and flow conditions.

\end{abstract}

\dates{This manuscript was compiled on \today}
\doi{\url{www.pnas.org/cgi/doi/10.1073/pnas.XXXXXXXXXX}}

\begin{document}

\maketitle
\thispagestyle{firststyle}
\ifthenelse{\boolean{shortarticle}}{\ifthenelse{\boolean{singlecolumn}}{\abscontentformatted}{\abscontent}}{}

{\replyOne The prediction of the statistical properties of turbulent flows} is critical for engineering (cars to nuclear reactors), science (ocean dynamics to astrophysics) and government policy (climate and weather forecasting). Over the last sixty years we have increasingly relied for such predictions on simulations based on the numerical integration of the Navier-Stokes equations. Today we can perform simulations using trillions of computational elements and resolve flow phenomena at unprecedented detail. However, despite the ever increasing availability of computing resources, most simulations of turbulent flows require the adoption of models to account for the spatio-temporal scales that cannot be resolved. Over the last few decades, the development of turbulence models has been the subject of intense investigations that have relied on physical insight and engineering intuition. Recent advances in machine learning and in the availability of data have offered new perspectives (and hope) in developing data-driven turbulence models.
{\replyThr The study of turbulent flows is rooted in the seminal works of Kolmogorov on statistical analysis~\citep{Kolmogorov1941}.} %Li1997
These flows are characterized by vortical structures, and their interactions, exhibiting a broad spectrum of spatio-temporal scales~\citep{taylor1935, pope2001}. %Cardesa2017 
At one end of the spectrum we encounter the integral scales, which depend on the specific forcing, flow geometry, or boundary conditions. At the other end are the Kolmogorov scales at which turbulent kinetic energy is dissipated. The handling of these turbulent scales provides a classification of turbulence simulations: Direct Numerical Simulations (DNS), which use a sufficient number of computational elements to represent all scales of the flow field, and simulations using turbulence models where the equations are solved in relatively few computational elements and the non-resolved terms are described by closure models.
{\replyOne While the flow structures at Kolmogorov scales are statistically homogeneous and dissipate energy, most of the computational effort of DNS~\citep{Moin1998} is spent in attempting to fully resolve them.}
DNS~\citep{moser1999} have provided us with unique insight into the physics of turbulence that can lead in turn to effective turbulence modeling. However, it is well understood that for the foreseeable future DNS will not be feasible at resolutions necessary for engineering applications. In the development of turbulence models~\citep{Durbin2018} two techniques have been dominant: Reynolds Averaged Navier-Stokes and Large-eddy Simulations (LES)~\citep{leonard1974} in which only the large scale unsteady physics are explicitly computed whereas the sub grid-scale (SGS), unresolved, physics are modeled.
In LES, classic approaches to the explicit modeling of SGS stresses include the standard~\citep{smagorinsky1963} and the dynamic Smagorinsky model~\citep{germano1991, lilly1992}. In the last 50 years SGS models have been constructed using physical insight, numerical approximations and often problem-specific intuition. {\replyTwo The first efforts to develop  models for turbulent flows using machine learning~\citep{lee1997,milano2002} were hindered by the available computing power and the convergence of the training algorithms. Recent advances in hardware and algorithms have fueled a broad interest in the development of data-driven turbulence models~\citep{duraisamy2019}.}

\begin{figure*}[b!] \centering
	\includegraphics[trim={0 0 0 0}, clip, width=.8\textwidth]{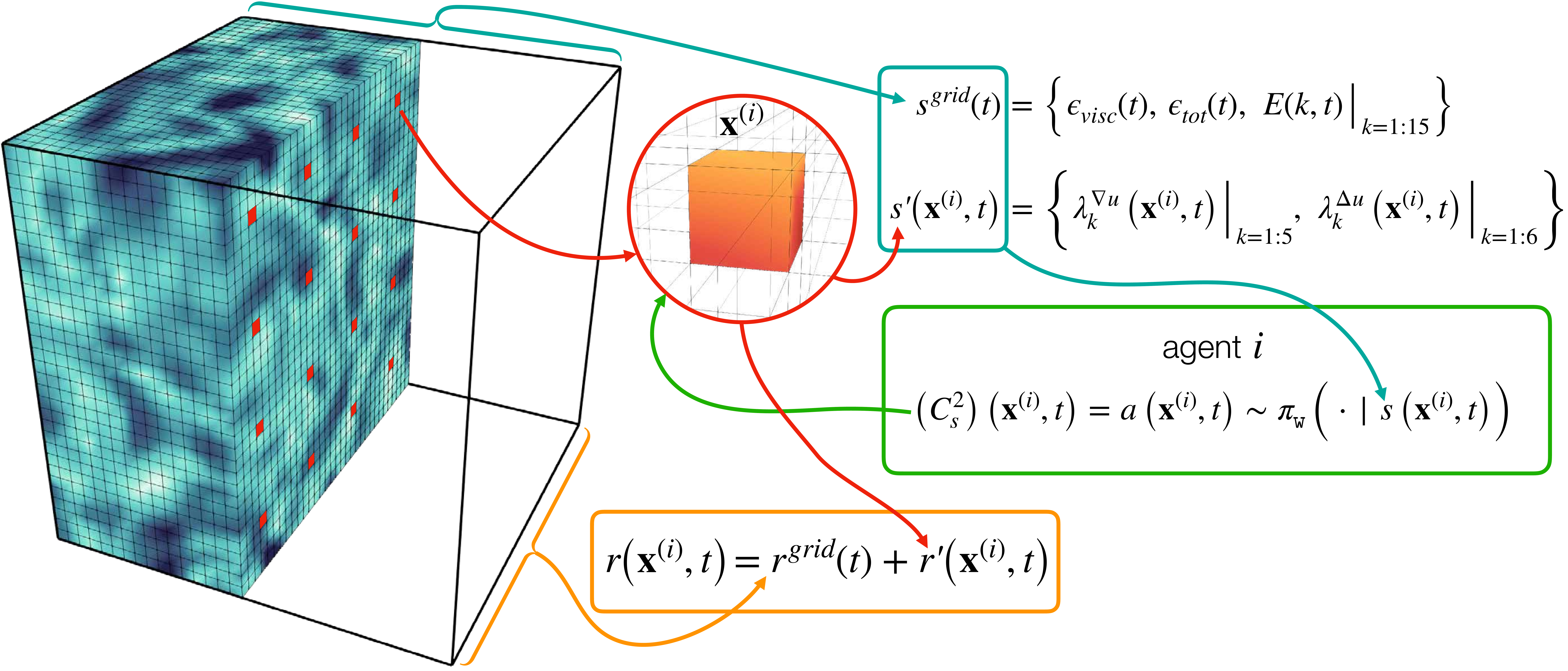}
	\caption{\textbf{Schematic of the integration of MARL with the flow solver.} The agents (located at the red blocks) compute the SGS dissipation coefficient $C_s^2$ for each grid-point of the simulation by sampling a shared control policy $\pi_{\mathtt w}(\cdot|\mathbf{s}(\mathbf{x}, t))$.
	The state $\mathbf{s}(\mathbf{x}, t)$ of agent $i$ at time $t$ is defined by both local variables (i.e. $\mathbf{s}'(\mathbf{x}, t)$, which includes the invariants of the gradient $\lambda_k^{\nabla \mathbf{u}}$ and Hessian $\lambda_k^{\Delta \mathbf{u}}$ of the velocity field computed at the agents' location $\mathbf{x}^{(i)}$), and global variables (i.e. $\mathbf{s}^{grid}(t)$, composed by the energy spectrum up to the Nyquist frequency, the rate viscous dissipation $\epsilon_{visc}$ and the total dissipation rate $\epsilon_{tot}$). Depending on the accuracy of the resulting LES simulation, the agents receive a scalar reward $r(\mathbf{x}, t)$ which can be similarly defined from both local and global components.
	} \label{fig:rl loop}
\end{figure*}

To date, to the best of our knowledge, all data-driven turbulence closure models are based on supervised learning.
In LES, early approaches~\citep{sarghini2003} trained a NN to emulate and speed-up a conventional, but computationally expensive, SGS model.
More recently, data-driven SGS models have been trained by supervised learning to predict the ``perfect'' SGS terms computed from filtered DNS data~\citep{gamahara2017, xie2019}. Variants include deriving the target SGS term from optimal estimator theory~\citep{vollant2017} and reconstructing the SGS velocity field as a deconvolution operation, or inverse filtering~\citep{hickel2004,maulik2017}.
In supervised learning, the parameters of the NN are commonly derived by minimizing the model prediction error via a gradient descent algorithm. As the error is required to be differentiable with respect to the model parameters, and due to the computational challenge of obtaining chain-derivatives through a flow solver~\citep{sirignano2019}, supervised learning approaches often rely on one-step target values for the model (e.g. SGS stresses computed from filtered DNS). Such \emph{a-priori} testing measures the accuracy of the  derived model in predicting the target values from a database of reference simulations, typically obtained via DNS. After training, \emph{a-posteriori} testing is performed by integrating in time the flow equations along with the learned closure and comparing the obtained statistical quantities to those from DNS or other references.
We remark that in the case of a single-step cost function, the resultant NN model is not trained to compensate for the evolution of discrepancies between DNS and LES data and the compounding errors. This critical issue of SGS models derived by supervised learning has been exposed by studies that perform \emph{a posteriori} testing~\citep{wu2018}. \emph{A-priori} perfect SGS models may be structurally unstable, accumulate high-spatial frequency errors and diverge from the original trajectory under small perturbations~\citep{nadiga2007, beck2019}. Moreover, models trained to reproduce a particular quantity of interest may worsen the accuracy of other physical quantities~\cite{gamahara2017}.

In this work, we address these challenges by introducing Reinforcement Learning (RL) as a framework for the automated discovery of closure models for non-linear conservation laws.
%and the construction of SGS models for LES.
Specifically, we analyze the potential of RL to control under-resolved simulations (LES) of isotropic turbulence by locally adapting the coefficients of the eddy viscosity closure model, with the objective of accurately reproducing the energy spectrum predicted by DNS.
Two characteristics of RL make it particularly suited to the task.
First, RL casts  the closure problem in terms of the actions of an agent that learns to optimize their long-term consequences on the environment.
In RL, training is not performed on a database of reference data, but by integrating in time the parametric model.
Consequently, the RL framework overcomes the above mentioned  distinction between \emph{a priori} and \emph{a posteriori} evaluation and accounts for compounding modeling errors. Moreover, the performance of a RL strategy is not measured by a differentiable objective function but by a cumulative reward. Supervised learning approaches that train a model to recover SGS quantities computed from filtered DNS simulations necessitate the computational capabilities of fully resolving the flow simulation. This is not required in RL, as the reward can be a measure of the similarity between the statistics of a quantity of interest produced by the model and reference data, which may even be obtained from experiments.
Finally, we remark that the proposed RL framework is not restricted to simulations of the Navier-Stokes equations and is readily adaptable to other non-linear conservation laws.

\section*{Multi-agent RL for sub-grid scale modeling}
{\replyTwo RL is a computational framework for control problems~\citep{sutton2018} which implies goal-directed interactions of an agent with its environment. RL is at the core of some of the seminal results of machine learning, in applications including games \citep{mnih2015,silver2016} and robotics \citep{levine2016,akkaya2019}.  
In RL the agent performs \emph{actions} that affect its environment. The agent's actions are contingent on its \emph{state} and the performance is measured via scalar \emph{reward} functions. By acquiring experience, the agent learns a policy ($\pi(a|s)$) from which it samples actions that maximize the long term, cumulative rewards.}
In recent years, RL has been making inroads in the field of flow control~\citep{garnier2019}. By interacting with the flow field, agents trained through RL were able to gather relevant information and optimize their decision process to perform collective swimming~\citep{gazzola2014}, soar~\citep{reddy2016}, minimize their drag~\citep{novati2017,verma2018}, delay the onset of instabilities~\citep{belus2019}, or reach a target location~\citep{biferale2019, novati2019}.

{\replyTwo A key aspect of RL is the representation of the policy function.
Deep RL algorithms train NN to represent the policy ($\pi_{\mathtt w}(\cdot|\mathbf{s})$, with parameters  $\mathtt w$), bypassing the need for tabular or expertly designed state representations~\citep{franccois2018}.} In the present study, a policy-network is used to sample the dissipation coefficient $C^2_s$ of the Smagorinsky SGS model.
We emphasize that the interface of RL with  the flow solver has considerable effect on the computational efficiency of the resulting model. As an example, following the common practice in video games~\citep{mnih2015}, the state $\mathbf{s}$ of the agent could be defined as the full three-dimensional flow field at a given time-step and the action as the SGS closure for all grid-points. With such an architecture, the dimensionality of both state and action spaces would scale with the number of degrees of freedom of the simulation. As a consequence, the closure model would be mesh-size dependent, would involve training a very large NN, and the memory needed to store the experiences of the agent would be prohibitively large.
{\replyTwo Here, we overcome these issues through multi-agent reinforcement learning (MARL).

In MARL, the $N_{agents}$ agents are dispersed in the simulation domain (Fig.~\ref{fig:rl loop}). Each agent ($i$, with spatial coordinate $\bm{x}^{(i)}$) performs a localized action based on information about the state of the flow field $s(\bm{x}^{(i)}, t) \in \mathbb{R}^{dim_S}$, which is encoded by a small set of local and global variables.
We embed tensorial invariance into the NN inputs~\citep{ling2016} by selecting as local variables of the state vector the 5 invariants~\citep{pope1975} of the gradient ($\lambda_k^{\nabla \mathbf{u}}$) and the 6 invariants of the Hessian of the velocity field ($\lambda_k^{\Delta \mathbf{u}}$). These are computed at the agents' location and non-dimensionalized with $K/\epsilon$. The global components of the state are the modes of the energy spectrum up to the training grid's Nyquist frequency $N_{nyquist}$ (non-dimensionalized with $u_\eta$), the rate of viscous dissipation ($\epsilon_{visc} / \epsilon$) and the total dissipation ($\epsilon_{tot} / \epsilon$) relative to the turbulent energy injection rate.
{\replyThr The training phase is performed on a Cartesian mesh of size $N=32^3$, with a pressure-projection scheme, second-order discretization of the spatial derivatives, and second-order explicit Runge-Kutta time integration. Because $N_{Nyquist} = 15$, we have state dimensionality $dim_S = 28$. These are far fewer variables than would be required by encoding as a state the entire velocity field of the flow ($dim_S = 3 \cdot 32^3$).}

MARL advances in turns by updating the dissipation coefficients $C^2_s(\bm{x}, t)$ for the whole flow and integrating the LES in time for $\Delta t_{RL}$, until $t=T_{end}$ or any numerical instability arises.
At the start of every turn, each agent $i$ measures its state and selects an action $a(\bm{x}^{(i)}, t) \in \mathbb{R}$ by sampling a Gaussian policy: $a(\bm{x}^{(i)}, t) \sim \pi_{\mathtt w}(\cdot \, | \, s(\bm{x}^{(i)}, t)) \equiv \mathcal{N}[\mu_{\mathtt w}\left(s(\bm{x}^{(i)}, t)\right), \, \sigma^2_{\mathtt w}\left(s(\bm{x}^{(i)}, t)\right)]$.
The actions are used to compute $C^2_s(\bm{x}, t)$ for each grid point by linear interpolation:
\begin{equation}
	C^2_s(\bm{x}, t) = \sum_{i=1}^{N_{agents}}  a(\bm{x}^{(i)}, t)  \prod_{j=1}^3 \max \left\{ 1 {-} \frac{|{x}_j {-} {x}^{(i)}_j|}{\Delta_{agents}}, ~0 \right\},
\end{equation}
where ${x}^{(i)}_j$ is the $j$-th Cartesian component of the position vector of agent $i$, and $\Delta_{agents} = 2 \pi / \sqrt[3]{N_{agents}}$ is the distance between agents. If $N_{agents} = N$, no interpolation is required.

Increasing $N_{agents}$ improves the adaptability of MARL to localized flow features. However, the actions of other agents are confounding factors that may increases the update variance~\citep{bucsoniu2010}. For example, if the $C_s^2$ coefficient selected by one agent causes numerical instability, all agents would receive negative feedback, regardless of their choices.} These challenges are addressed by performing policy optimization with the  Remember and Forget Experience Replay algorithm (ReF-ER)~\cite{novati2018}.
Three features of {ReF-ER} make it particularly suitable for MARL and the present task: First, as it relies on Experience Replay (ER)~\citep{lin1992}, it reuses experiences over multiple policy iterations and increases the accuracy of gradient updates by computing expectations from uncorrelated experiences.
Moreover, {ReF-ER} is inherently stable and has been shown to reach, and even surpass, the performence of state of the art RL algorithms and  optimal control~\citep{novati2019} methods on several benchmark problems. 
Finally, and crucially for MARL, {ReF-ER} explicitly controls the pace of policy changes. We found that {ReF-ER}, with strict constraints on the policy updates from individual experiences, is necessary to stabilize training and compensate for the imprecision of the single-agent update rules.
For a thorough description of ReF-ER see the Methods and the original paper~\citep{novati2018}.

MARL finds the parameters ${\mathtt w}$ that maximize the expected sum of rewards over each simulation $J({\mathtt w}) = \mathop\mathbb{E}_{\pi_{\mathtt w}} \left[ \sum_{t=1}^{T_{end}} r_t\right]$. In the present study, the parameters ${\mathtt w}$ are shared by all agents. Their aggregate experiences are collected in a shared dataset and used to computes updates according to ReF-ER (Fig.~\ref{fig:rl hpc}).
We define reward functions with the objective of obtaining policies $\pi_{\mathtt w}$ that yield stable LES and exhibit statistical properties that closely match those of DNS.
We find that the distribution of energy spectrum ${E}(k)$ of isotropic turbulence computed by DNS is well approximated by a log-normal distribution (see Methods and Fig.~\ref{fig:dns all}) such that $\log E^{Re_\lambda}_{DNS} \sim \mathcal{N}\left(\mu^{Re_\lambda}_{DNS} ,~ \Sigma^{Re_\lambda}_{DNS} \right)$, where $\mu^{Re_\lambda}_{DNS}$ is the average log-energy spectrum for a given $Re_\lambda$ and $\Sigma^{Re_\lambda}_{DNS}$ is its covariance matrix. When comparing SGS models and formulating objective functions, we rely on a regularized log-likelihood:
\begin{equation}
	\widetilde{LL}(E^{Re_\lambda}_{LES} | E^{Re_\lambda}_{DNS}) = \log \mathcal P (E^{Re_\lambda}_{LES} | E^{Re_\lambda}_{DNS}) / N_{nyquist}. \label{eq:ref loglikelihood}
\end{equation}
Here the probability metric is
\begin{multline}
	\mathcal P (E^{Re_\lambda}_{LES} | E^{Re_\lambda}_{DNS}) \propto \exp \bigg[ - \frac{1}{2}
	\left(\log E^{Re_\lambda}_{LES} - \bar{\mu}^{Re_\lambda}_{DNS} \right)^T \\ \left( \bar{\Sigma}^{Re_\lambda}_{DNS}\right)^{-1}
	\left(\log E^{Re_\lambda}_{LES} - \bar{\mu}^{Re_\lambda}_{DNS} \right) \bigg] \label{eq:ref probability}
\end{multline}
with $E_{LES}$ the LES energy spectrum, $\bar{\mu}^{Re_\lambda}_{DNS}$ and $\bar{\Sigma}^{Re_\lambda}_{DNS}$ the target statistics up to $N_{nyquist}$.

We consider two SGS models derived from MARL, each corresponding to a reward function of the form $r(\bm{x}^{(i)}, t)$.
The first ($\pi^{G}_{\tt w}$) is defined by rewards $r^G(\bm{x}^{(i)}, t)$ based on the error in the Germano identity~\citep{germano1991} which states that the sum of resolved and modeled contributions to the SGS stress tensor should be independent of LES resolution.
The second ($\pi^{LL}_{\tt w}$) is defined by rewards $r^{LL}(\bm{x}^{(i)}, t)$ which strongly penalize discrepancies from the target energy spectra. 
While $r^{G}$ is computed locally for each agent, $r^{LL}$ equal for all agents.
We remark that the target statistics involve spatial and temporal averages and can be computed from a limited number of DNS, which for this study are four orders of magnitude more computationally expensive than LES.

%WAS Results
\section*{LES modeling of DNS} \label{sec:results}

\begin{figure*}[t!]
	\includegraphics[width=\textwidth]{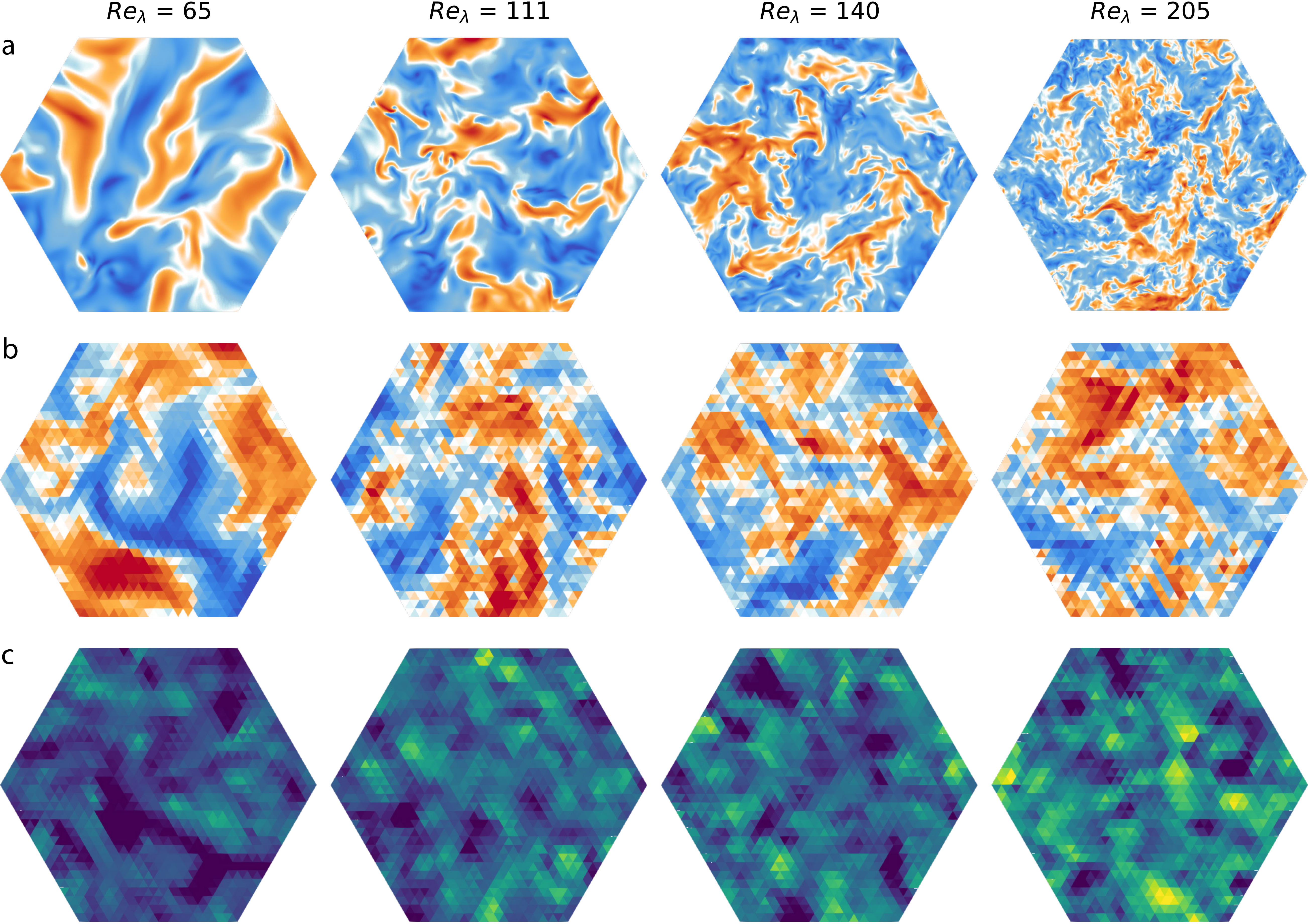}
	\caption{\textbf{Visualizations of simulations of isotropic turbulence.}
		\textbf{a}, Representative contours of momentum flux across a diagonal slice ($x+y+z=0$) of the cubical domain ($\bm u \cdot \bm n$, blue and orange indicate negative and positive fluxes) for DNS of isotropic turbulence with resolution $1024^3$,
		\textbf{b}, for LES with resolution $32^3$ and SGS modeling with a RL policy trained for $r^{LL}$.
		\textbf{c}, Contours of the Smagorinsky coefficient $C_s^2$ across the same diagonal slice of the LES (blue and yellow indicate low and high values respectively).
	} \label{fig:contours}
\end{figure*}
The Taylor-Reynolds number ($Re_\lambda$) characterizes the breadth of the spectrum of vortical structures present in a isotropic turbulent flow~\cite{taylor1935,pope2001}.
Figure~\ref{fig:contours} illustrates the challenge in developing a reliable SGS model for a wide range of $Re_\lambda$ and for a severely under-resolved grid. Only the large eddies are resolved and for the lower Reynolds numbers (e.g. $Re_\lambda = 65$) the SGS model is barely able to represent the flow features of DNS.
{\replyTwo For Reynolds numbers beyond $Re_\lambda{=}111$
the fluctuating vortical structures that characterize the turbulent flow field occur at length-scales that are much smaller than the LES grid-size. 
As a consequence, an increasing portion of energy dissipation is due to SGS effects, which leads to instability if these are not accurately modeled.}

\begin{figure*}[t!]
	\centering \includegraphics[trim={0 0 0 0},clip,width=\linewidth]{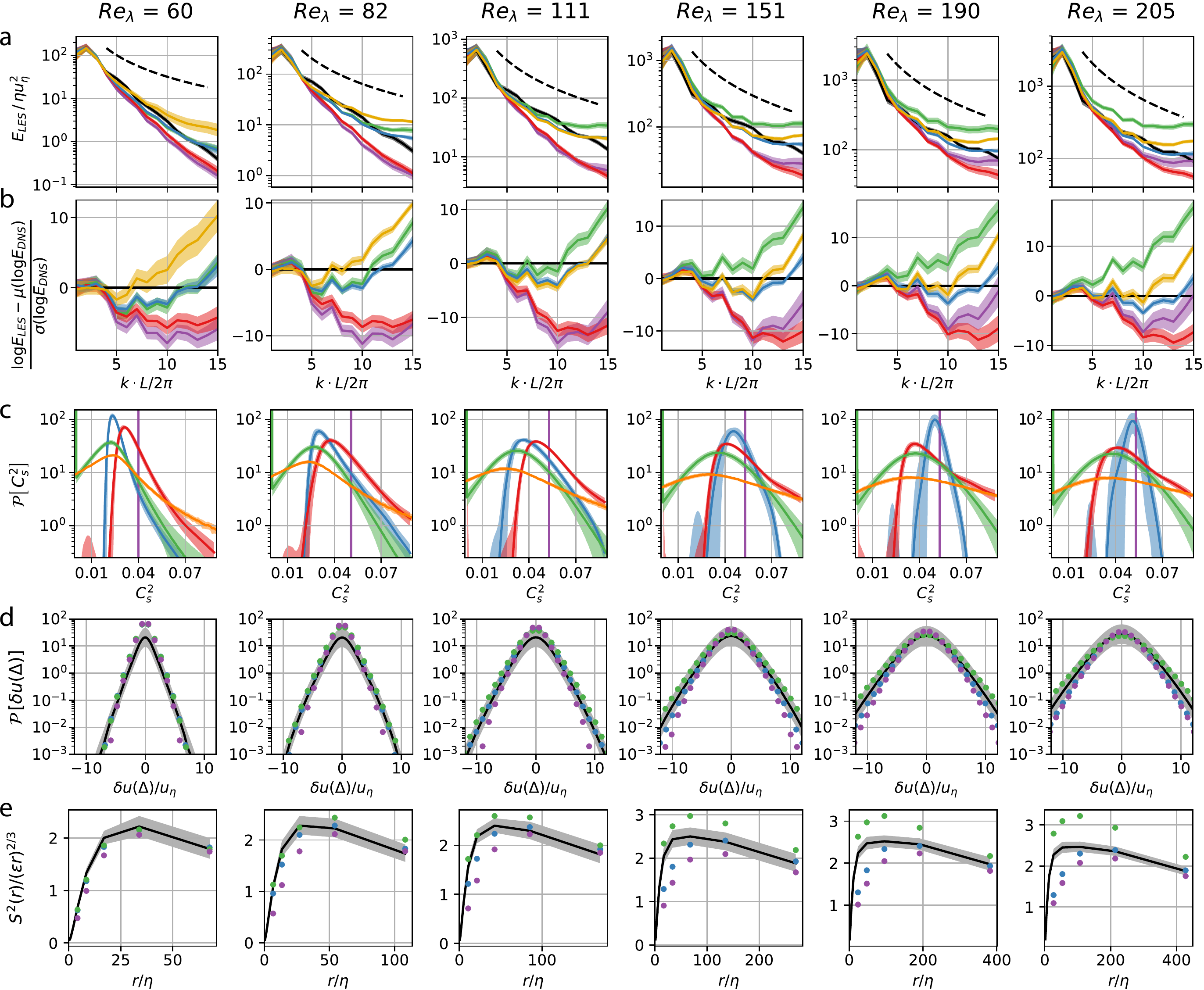}
	\caption{\textbf{
	{\replyOne Comparison of SGS models for values of $Re_\lambda$ that were \emph{not} included during the MARL training.}} \textbf{a}, Energy spectra for ({\rule[0.5ex]{1.5em}{1pt}})~DNS,
	({\color{RLtrainSSM}  \rule[0.5ex]{1.5em}{1pt}})~SSM,
	({\color{RLtrainDSM}  \rule[0.5ex]{1.5em}{1pt}})~DSM,
	({\color{RLtrainLL}   \rule[0.5ex]{1.5em}{1pt}})~MARL policy $\pi^{LL}_{\tt w}$, 
	({\color{RLtrainGERM} \rule[0.5ex]{1.5em}{1pt}})~MARL policy $\pi^{G}_{\tt w}$, and ({\color{RLtrainSINGLE}   \rule[0.5ex]{1.5em}{1pt}})~MARL policy $\pi^{LL, 111}_{\tt w}$ trained exclusively from data for $Re_\lambda = 111$, compared to ({\rule[0.5ex]{0.5em}{1pt}\,\rule[0.5ex]{0.5em}{1pt}}) the -5/3 Kolmogorov scaling.
	\textbf{b}, Log-energy spectra normalized by the DNS mean and the standard deviation.
	\textbf{c}, Empirical probability distributions of the Smagorinsky model coefficient $C_s^2$ for the first four models (same colors) compared to ({\color{RLtrainGRU} \rule[0.5ex]{1.5em}{1pt}}) the SGS dissipation coefficient computed from DNS filtered to LES resolution. \textbf{d}, distribution of longitudinal velocity increments $\delta u(r) = \left[\mathbf{u}(\mathbf{x} + \mathbf{r}) - \mathbf{u}(\mathbf{x})\right] \cdot \hat{\mathbf{r}}$ for $r$ equal to the LES grid size $\Delta$ and, \textbf{e}, second-order velocity structure function $S^2(r) = \langle \delta u(r)^2 \rangle$  for ({\color{RLtrainLL} $\bullet$})~$\pi^{LL}_{\tt w}$, ({\color{RLtrainDSM} $\bullet$})~DSM, and ({\color{RLtrainSSM} $\bullet$})~SSM compared to DNS. In all cases, lines represent averages and contours represent intervals of one standard deviation.} \label{fig:RL convergence}
\end{figure*}

A natural way to assess whether the turbulence model accurately reproduces the energy transfer to SGS motions is through the energy spectra.
In Fig.~\ref{fig:RL convergence}\textbf{a}, we compare the time-averaged spectra obtained by DNS to those obtained by LES with several SGS closure models. More specifically, we show the first $N_{nyquist}$ modes of the energy spectra, and their normalization with the mean and standard deviation of the energy computed through DNS (Fig.~\ref{fig:RL convergence}\textbf{b}). This measure quantifies the contributions of individual modes to the objective log-likelihood (Eq.~\ref{eq:ref loglikelihood}). A perfect SGS model would produce a spectrum with time-averaged $E^{Re_\lambda}_{LES}(k)$ with the same statistics as that from a DNS. We consider the two classical approaches, the standard Smagorinsky model (SSM-with an empirically tuned constant dissipation coefficient $C_s$), and the dynamic Smagorinsky model (DSM- with an adaptive coefficient derived from the Germano identity). The two models serve as a reference for the accuracy of SGS models derived through MARL, identified by the policies $\pi_{\mathtt w}^{LL}$ and $\pi_{\mathtt w}^{G}$. We remark that the models are evaluated in flows with  Reynolds numbers which were not presented during training. The amount of SGS dissipation, as well as the the numerical scales of the flow quantities, and of the RL state components, vary with $Re_\lambda$. Therefore, the results for $Re_\lambda =82$, $111$, and $151$ measure the MARL model accuracy for dynamical scales that are interposed with the training ones, while the results for $Re_\lambda =60$, $190$ and $205$ measures the ability of the MARL models to generalize beyond the training experiences.

{\replyThr The MARL model trained to satisfy the Germano identity highlights an important consequence of RL maximizing long term rewards. DSM, which minimizes the instantaneous Germano-error, exhibits growing energy build-up at high-frequencies, which causes numerical instability at higher $Re_\lambda$. In fact, the Germano identity is not expected to be accurate for severely under-resolved LES. Conversely, $\pi_{\mathtt w}^{G}$, which minimizes the error over all future steps, over-estimates the dissipation coefficient, smoothing the velocity field, and making it easier for future actions to satisfy the Germano identity. This can be otherwise observed by Fig.~\ref{fig:RL convergence}\textbf{c}, which shows the empirical distribution of Smagorinsky coefficients chosen by the SGS models. While outwardly DSM and $\pi_{\mathtt w}^{G}$ minimize the same relation, $\pi_{\mathtt w}^{G}$ introduces much more artificial viscosity.}

The policy $\pi_{\mathtt w}^{LL}$, which directly maximizes the similarity between quantity of interest (i.e. energy spectra) obtained by MARL and those of DNS, produces the SGS model of the highest quality.
While its accuracy is similar to that of DSM for lower values of $Re_\lambda$, $\pi^{LL}_{\mathtt w}$ avoids energy build-up and remains stable up to $Re_\lambda = 205$, well beyond the maximum training $Re_\lambda=163$. Higher Reynolds numbers were not tested as they would have required increased spatial and temporal resolution to carry out accurate DNS, with prohibitive computational cost.
We evaluate the difficulty of generalizing beyond the training data by comparing $\pi^{LL}_{\mathtt w}$ to a policy fitted exclusively for $Re_\lambda = 111$ ($\pi^{LL,~111}_{\mathtt w}$). Fig.~\ref{fig:RL convergence}\textbf{b} shows the specialized policy to have comparable accuracy at $Re_\lambda = 111$, but becomes rapidly invalid when varying the dynamical scales. This result supports that data-driven SGS models should be trained on varied flow conditions rather than with a training set produced by a single simulation.

\begin{figure*}
    \includegraphics[width=\textwidth]{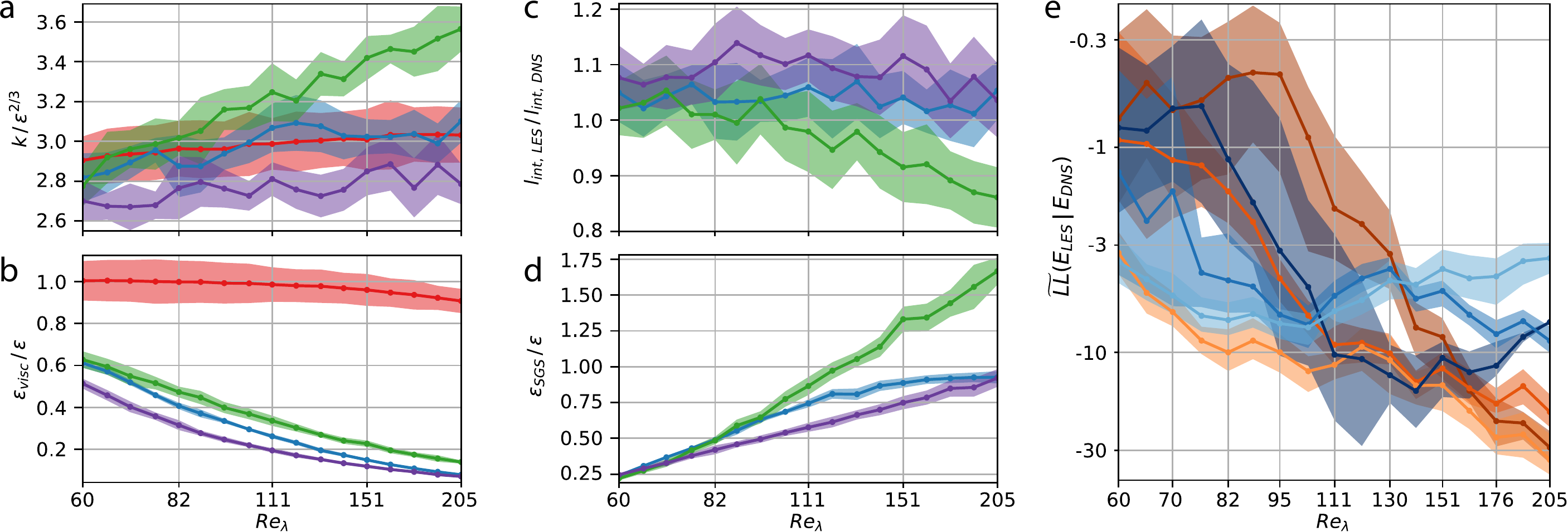}
	\caption{\textbf{Measurements of the reliability of the MARL model beyond the training objective.} Integral properties of LES: \textbf{a}, turbulent kinetic energy, \textbf{b}, integral length scale, \textbf{c}, ratio of viscous dissipation to energy injection, and \textbf{d}, ratio of SGS dissipation to energy injection. The remaining component of energy dissipation is due to numerical discretization.
	({\color{RLtrainSSM}  \rlap{$\:\bullet$}\rule[0.4ex]{1.2em}{1pt}})~SSM,
		({\color{RLtrainDSM}  \rlap{$\:\bullet$}\rule[0.4ex]{1.2em}{1pt}})~DSM,
		({\color{RLtrainLL}   \rlap{$\:\bullet$}\rule[0.4ex]{1.2em}{1pt}})~MARL policy $\pi_{\mathtt w}^{LL}$, and
		({\color{RLtrainGERM} \rlap{$\:\bullet$}\rule[0.4ex]{1.2em}{1pt}})~DNS simulation when applicable.
	\textbf{e}, Accuracy across grid-resolutions measured as log-likelihood of LES spectra with respect to DNS statistics (Eq.~\ref{eq:ref loglikelihood}) for DSM at resolutions ({\color[HTML]{fd8d3c} \rlap{$\:\bullet$}\rule[0.4ex]{1.2em}{1pt}})~$N{=}32^3$, ({\color[HTML]{e6550d} \rlap{$\:\bullet$}\rule[0.4ex]{1.2em}{1pt}})~$N{=}64^3$, ({\color[HTML]{a63603} \rlap{$\:\bullet$}\rule[0.4ex]{1.2em}{1pt}})~$N{=}128^3$ and MARL policy $\pi_{\mathtt w}^{LL}$ at resolutions ({\color[HTML]{6baed6} \rlap{$\:\bullet$}\rule[0.4ex]{1.2em}{1pt}})~$N{=}32^3$, ({\color[HTML]{2171b5} \rlap{$\:\bullet$}\rule[0.4ex]{1.2em}{1pt}})~$N{=}64^3$, ({\color[HTML]{08306b} \rlap{$\:\bullet$}\rule[0.4ex]{1.2em}{1pt}})~$N{=}128^3$.
	In all cases, data points correspond the temporal averages, and contours denote intervals of one standard deviation.
	} \label{fig:generalization}
	%\rule[0.5ex]{1.0em}{1pt}
\end{figure*}

From Fig.~\ref{fig:RL convergence}\textbf{c}, we observe that $\pi_{\mathtt w}^{LL}$ achieves its accuracy by producing a narrower distribution of $C_s^2$.
In this respect, $\pi_{\mathtt w}^{LL}$ stands in contrast to a model trained by supervised learning to reproduce the SGS stresses computed from filtered DNS.
By filtering the DNS results to the same resolution as the LES, thus isolating the unresolved scales, we emulate the distribution of $C_s^2$ that would be produced by a SGS model trained by supervised learning.
We find that such model would have lower SGS dissipation than both DSM and $\pi_{\mathtt w}^{LL}$, suggesting that, with the present numerical discretization schemes, it would produce numerically unstable LES.
In fact, it is well known that the numerical discretization errors due to the coarse LES resolution, which may not be anticipated when using filtered DNS as training data, may be comparable or larger than the modeling errors (i.e. the errors due to the SGS model)~\citep{meyers2003}.
This further highlights the unique ability of MARL to systematically optimize high-level objectives, such as matching the statistics of DNS, and suggests its potential in deriving data-driven closure equations.

\section*{Generalization beyond the training objective} \label{sec:generalize}

{\replyThr The energy spectrum is just one of many statistical quantities that a physically sound LES should accurately reproduce. In fact, MARL in order to maximize its objective function may inadvertently sacrifice the physical soundness of other quantities.
Fig.~\ref{fig:RL convergence}\textbf{d} shows the distribution of relative velocity between two adjacent points of the LES grid. We see that the tails of the distribution are tapered by $\pi_{\mathtt w}^{LL}$ and SSM in order to maintain stability. }
{\replyOne Fig.~\ref{fig:RL convergence}\textbf{e} shows the second-order velocity structure function $S^2(r)$, which is the covariance of the velocity between two points separated by a distance $r$.
According to Kolmogorov's hypothesis, for $r \gg \eta$, the structure function should have scaling behavior $S^2(r) \propto r^{\zeta(p)}$ with coefficient depending only on the energy flux $\epsilon$~\citep{Kolmogorov1941,pope2001}. As expected from DNS, the scaling of $S^2(r)$ approaches a constant value regardless of $Re_\lambda$. }
In Fig.~\ref{fig:generalization}\textbf{a-d} we compare the total kinetic energy (TKE), the characteristic length scale of the largest eddies ($l_\text{int}$), and dissipation rates among LES models and DNS.
While in DNS energy is dissipated entirely by viscosity (and if under-resolved by numerical diffusion), in LES the bulk of viscous effects occur at length-scales below the grid size, especially at high $Re_\lambda$.
We find that for $Re_\lambda = 205$ the stable SGS models dissipate approximately 10 times more energy than viscous dissipation, which underlines the crucial role of turbulence modeling.
Up to the point of instability at $Re_\lambda \approx 100$, DSM yields a good estimate for TKE and $l_\text{int}$. Beyond that value, the artificial energy created by numerical instabilities causes the SGS dissipation to increase past the energy injection rate $\epsilon$.
Despite these quantities not being directly included in the rewards, they are all correctly recovered by the MARL model $\pi_{\mathtt w}^{LL}$.

Finally, we evaluate MARL across grid resolutions. Because of the design of the MARL framework, the policy $\pi_{\mathtt w}^{LL}$, trained for a single grid-size $N=32^3$, is valid as long as there exist at least 15 modes of the energy spectrum.
In Fig.~\ref{fig:generalization}\textbf{e}~we compare the log-likelihood of DSM and MARL models given the DNS statistics for $N=32^3$, $64^3$ and $128^3$. Accordingly, we increase the number of agents per simulation by a factor of 8 and 64 to keep constant the density of agents in the grid. We remark that LES at finer resolutions more accurately represent large-scale statistics, but this is not reflected in the values of the log-likelihood. In fact, at finer resolutions more modes are included and $\widetilde{LL}$ is dominated by errors at the high frequencies. Moreover, because only the first 15 components of the spectrum are available to $\pi_{\mathtt w}^{LL}$, MARL agents were not trained to take into account higher energy modes. Finer resolutions are able to capture sharper velocity gradients not experienced during training.
As a consequence, $\pi_{\mathtt w}^{LL}$ was found to be more diffusive than DSM at the higher frequencies. Nevertheless, the SGS model derived by MARL remains stable throughout the evaluation and markedly more accurate than DSM, especially at higher values of $Re_\lambda$.

\section*{Discussion}
This paper introduces multi-agent RL (MARL) to automate the discovery  of closure models in simulations of turbulent flows.  We demonstrate the feasibility and potential of this approach on large-eddy simulations (LES) of forced isotropic turbulence. MARL develops the sub-grid scale (SGS) closure as a control policy enacted by cooperating agents. The agents are incorporated into the flow solver, observe local (e.g. invariants of the velocity gradient) as well as global (e.g. the energy spectrum) flow quantities, and accordingly compute SGS residual-stresses through the Smagorinsky~\citep{smagorinsky1963} formulation. The Remember and Forget Experience Replay (ReF-ER) method is instrumental for the present study by combining the sample-efficiency of ER and the stability of constrained policy updates~\citep{novati2018}.

We believe that the learning algorithms and results of the present study open new horizons for turbulence modeling efforts. RL maximizes high-level objectives computed from direct application of the learned model and produces SGS models that are stable under perturbation and resistant to compounding errors. Here, MARL minimizes the discrepancies between the energy spectra of LES and that computed from orders of magnitude more computationally expensive, fully resolved simulations (DNS). Access to DNS targets allowed us to vary systematically $Re_\lambda$ and analyze the accuracy and generality of the trained model with respect to multiple quantities of interest.

New questions emerge from integrating deep learning and turbulence modeling.
In the present study, the control policies trained by MARL (e.g. $\pi_{\mathtt w}^{LL}$) are functions with 28-dimensional input and 6'211 parameters which encode the complex correlations between input and eddy-viscosity coefficient. {\replyThr In fact, the present policies may not be readily transferrable to another flow solver which uses a different numerical discretization.} While machine learning approaches can be faulted for the lack of generality guarantees and for the difficulty of interpreting the trained model, we envision that sparse RL methods could enable the analysis of causal processes in turbulent energy dissipation and the distillation of mechanistic models.
{\replyThr Moreover, the MARL framework, when applied to a new solver, will again learn to compensate for its modeling errors, a capability that may not be readily available to other closure models.
Finally, data-driven models, optimized for one type of reward may violate other physical constraints and invariants of the flow. Empirical results indicate that this issue does not affect the present results, but it remains an open question.} 

At the same time, the MARL framework opens many new directions for building upon the present results. MARL offers new perspectives on addressing classic challenges of LES, such as wall-layer modeling and inflow boundary conditions~\citep{zhiyin2015}. {\replyOne Moreover, the ability of performing DNS is not required to train RL models. Energy spectra, wall shear stresses or drag coefficients may be measured from experiments and used to train SGS models for LES, avoiding the computational expense of DNS.} {\replyThr In fact, many turbulent flows are not stationary, homogeneous or isotropic. It may be prohibitively expensive to measure statistical properties for such flows through multiple realizations of DNS. It is an open, and exciting, research direction to examine whether instantaneous, noisy, experimental measurements of multiple quantities of interest, translated into a reward function, may be used to successfully train SGS closures through RL.}

\matmethods{
\textbf{The Reinforcement Learning framework} \label{IRL}
RL algorithms advance by trial-and-error exploration and are known to require large quantities of interaction data, in this case acquired by performing thousands of LES with modest but non-negligible cost (which is orders of magnitude higher than the cost of ordinary differential equations or many video games). Therefore, the design of a successful RL approach must take into account the actual computational implementation. Here we rely on the open-source RL library \texttt{smarties}, which was designed to ease high-performance interoperability with existing simulation software.
\texttt{smarties} efficiently leverages the computing resources by separating the task of updating the policy parameters from the task of collecting interaction data (Fig.~\ref{fig:rl hpc}). The flow simulations are distributed across $N_{workers}$ computational nodes (``workers''). The workers collect, for each agent, experiences organized into episodes: $$\tau_{i} = \left\{s_t^{(i)},\,r_t^{(i)},\,\mu_t^{(i)},\,\sigma_t^{(i)},\,a_t^{(i)}\right\}_{t=0:T^{(i)}_{end}}$$ where $t$ tracks in-episode RL steps; $\mu_t^{(i)}$ and $\sigma_t^{(i)}$ are the statistics of the Gaussian policy used to sample $a_t^{(i)}$ with the policy parameters available to the worker at time step $t$ of the $i$-th episode, often termed ``behavior policy'' $\beta_t^{(i)} \equiv \mathcal{N}(\mu_t^{(i)},~\sigma_t^{(i)})$ in the off-policy RL literature. When a simulation concludes, the worker sends one episode per agent to the central learning process (``master'') and receives updated policy parameters. Therefore each simulation produces $N_{agents}$ episodes. The master stores the episodes into a Replay Memory (RM), which is sampled to update the policy parameters according to ReF-ER~\citep{novati2018}.

ReF-ER can be combined with many ER-based RL algorithms as it consists in a modification of the optimization objective. %For example, it has been applied to Q-learning, deterministic policy gradients \citep{lillicrap2015}, off-policy policy gradients \citep{wang2016}.
Here we employ {V-RACER}, a variant of off-policy policy optimization proposed in conjunction with {ReF-ER} which supports continuous state and action spaces. Because the cooperating agents do not explicitly coordinate their actions, the algorithm is unchanged from its initial publication.
{V-RACER} trains a Neural Network (NN) which, given input $s_t$, outputs the mean $\mu_{\mathtt w}\left(s_t\right)$ and standard deviation $\sigma_{\mathtt w}\left(s_t\right)$ of the policy $\pi_{\mathtt w}$, and a state-value estimate $v_{\mathtt w}(s_t)$. One gradient is defined per NN output.
The statistics $\mu_{\mathtt w}$ and $\sigma_{\mathtt w}$ are updated with the \emph{off-policy policy gradient} (\emph{off-PG})~\citep{degris2012}:
\begin{multline}
	{g}^\textrm{pol}({\mathtt w}) = \mathbb{E}\bigg[ \left. \left(\hat{q}_t - v_{\mathtt w}(s_t)\right) \frac{\pi_{\mathtt w}(a_t |  s_t)}{\mathcal{P}(a_t | \mu_t,\sigma_t)} \nabla_{\mathtt w} \log \pi_{\mathtt w}(a_t|s_t) \right| \\ 
	\{s_t,r_t,\mu_t,\sigma_t,a_t,\hat{q}_t\} \sim RM \bigg]. \label{eq:RMoffPG}
\end{multline}
Here $\mathcal{P}(a_t | \mu_t,\sigma_t)$ is the probability of sampling $a_t$ from a Gaussian distribution with statistics $\mu_t$ and $\sigma_t$, and $\hat{q}_t$ estimates the cumulative rewards by following the current policy from $(s_t,a_t)$ and is computed with the Retrace algorithm~\citep{munos2016}:
\begin{equation}
	\hat{q}_t = r_{t+1} + \gamma v_{\mathtt w}\left(s_{t+1}\right) + \gamma \min  \left\{1, \frac{\pi_{\mathtt w}(a_t | s_t)}{\mathcal{P}(a_t | \mu_t,\sigma_t)} \right\} \left[\hat{q}_{t+1} {-} v_{\mathtt w}\left(s_{t+1}\right)\right],~\label{eq:retrace}
\end{equation}
with $\gamma = 0.995$ the discount factor for rewards into the future.
Equation~\ref{eq:retrace} is computed via backward recursion when episodes are entered into the RM (note that $\hat{q}_{T_{end}} \equiv 0$), and iteratively updated as individual steps are sampled. Retrace is also used to derive the gradient for the state-value estimate:
\begin{multline}
	{g}^{val}({\mathtt w}) = \mathbb{E}\bigg[\left. \min \left\{1, \frac{\pi_{\mathtt w}(a_t | s_t)}{\mathcal{P}(a_t | \mu_t,\sigma_t)} \right\} \left(\hat{q}_t - v_{\mathtt w}(s_t)\right)  \right| \\ \{s_t,r_t,\mu_t,\sigma_t,a_t,\hat{q}_t\} \sim RM \bigg] \label{eq:RMoffVal}
\end{multline}
The \emph{off-PG} formalizes trial-and-error learning; it moves the policy to make actions with better-than-expected returns ($\hat{q}_t > v_{\mathtt w}(s_t)$) more likely, and those with worse outcomes ($\hat{q}_t < v_{\mathtt w}(s_t)$) less likely. Both Eq.~\ref{eq:RMoffPG} and Eq.~\ref{eq:RMoffVal} involve expectations over the empirical distribution of experiences contained in the RM, which are approximated by Monte Carlo sampling from the $N_{RM}$ most recent experiences $\hat{g}({\mathtt w}) = \sum\nolimits_{i=1}^B \hat{g}_i({\mathtt w})$, where $B$ the mini-batch size.
Owing to its use of ER and importance sampling, {V-RACER} and similar algorithms become unstable if the policy $\pi_{\mathtt w}$, and the distribution of states that would be visited by $\pi_{\mathtt w}$, diverges from the distribution of experiences in the RM. A practical reason for the instability may be the numerically vanishing or exploding importance weights $\pi_{\mathtt w}(a_t | s_t) / \mathcal{P}(a_t | \mu_t,\sigma_t)$. ReF-ER is an extended ER procedure which constrains policy changes and increases the accuracy of the gradient estimates by modifying the update rules of the RL algorithm:
\begin{equation}
	\hat{g}_t({\mathtt w}) \leftarrow \begin{cases}
		\beta  \hat{g}_t({\mathtt w}) \, {-}(1-\beta) g^D_t({\mathtt w})  & \textrm{if} ~\frac{1}{C}< \frac{\pi_{\mathtt w}(a_t | s_t) }{ \mathcal{P}(a_t | \mu_t,\sigma_t)}<C\\
		\qquad\quad{-}(1-\beta)  g^D_t({\mathtt w}) & \textrm{otherwise} 
	\end{cases} \label{eq:refer}
\end{equation}
here $g^D_t({\mathtt w}) = \nabla_{\mathtt w} D_{KL}\left(\pi_{\mathtt w}(\cdot | s_t) \,\|\, \mathcal{P}(\cdot | \mu_t,\sigma_t)\right)$ and $D_{KL}\left(P \,\|\, Q\right)$ is the Kullback-Leibler divergence measuring the distance between distributions $P$ and $Q$.
Equation~\ref{eq:refer} modifies the NN gradient by: 1) Rejecting samples whose importance weight is outside of a trust region determined by $C>1$. 2) Adding a penalization term to attract $\pi_{\mathtt w}(a_t | s_t)$ towards prior policies. The coefficient $\beta$ is iteratively updated to keep a constant fraction $D\in[0,1]$ of samples in the RM within the trust region:
\begin{equation}
	\beta \leftarrow \begin{cases}
		(1-\eta)\beta & \textrm{if}~n_{far} / N_{RM} > D \\
		\beta + (1-\eta)\beta & \textrm{otherwise}
	\end{cases} \label{eq:refer beta}
\end{equation}
Here $n_{far}/N_{RM}$ is the fraction of the RM with importance weights outside the trust region.

\begin{figure*} \centering
	\includegraphics[width=0.8\linewidth]{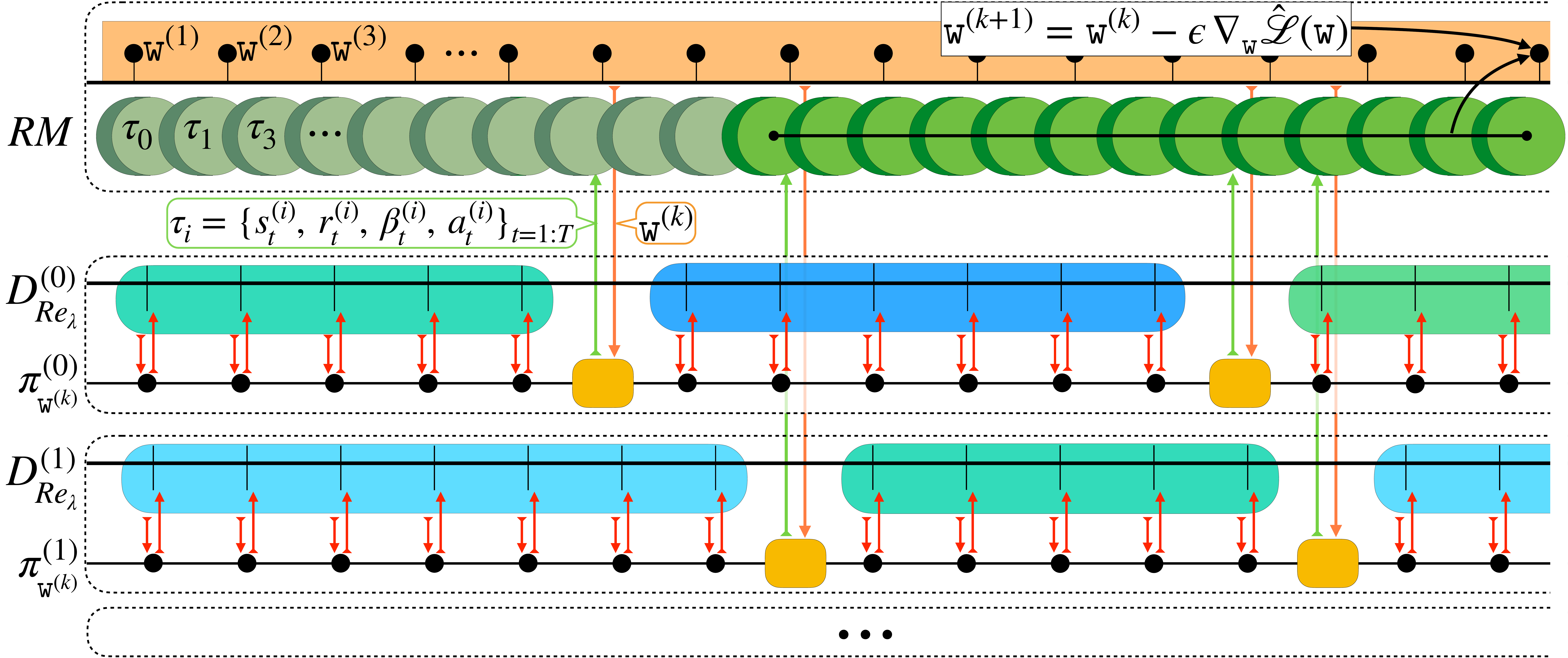}
	\caption{\textbf{Schematic description of the training procedure implemented with the \texttt{smarties} library.} Each dashed line represents a separate process. Multiple worker processes $j$ run LES for randomly sampled $Re_\lambda \in \left\{65,\,76,\,88,\,103,\,120,\,140,\,163\right\}$. The Reynolds number determines the dynamics ($D^{(j)}_{Re_\lambda}$) and the number of RL steps per simulation (i.e. the number of times the SGS dissipation coefficient is updated for the entire flow). At the end of each simulation, workers send full episodes $\tau_i$ (one per agent in the grid) and receive updated policy parameters. At the top, the diagram outlines  the two main tasks performed by the master process: (i) storing the $N$ most recently collected RL steps into a Replay Memory (RM) and (ii) sampling mini-batches from the RM in order to iteratively update the policy parameters $\mathtt w^{(k)}$ by gradient descent, advancing the update counter $k$. The policy is updated once every time any worker integrates its simulation over the RL step (i.e. one gradient step per SGS field update).}
	\label{fig:rl hpc}
\end{figure*}

\textbf{Overview of the training set-up} \label{RL:RLSM}
The two most notable hyper-parameters used in our description of the MARL setup are the actuation frequency (determined by $\Delta t_{RL}$) and the spatial resolution for the interpolation of the RL actions onto the grid (determined by $N_{agents}$). 
Both hyper-parameters serve the purpose of cutting down the amount of experiences collected during each simulation. The alternative would be to use the policy to compute $C_s^2$ for each grid-point of the domain and update its value on every simulation time-step. This would produce $\mathcal{O}(10^9)$ experiences per simulation and would make the temporal credit-assignment task (i.e. the RL objective of finding causal correlation between single actions and the observed reward) all the more difficult. The default values $\Delta t_{RL} = \tau_\eta / 8$ and $N_{agents} = 4^3$ reduce the number of experiences generated per simulation to $\mathcal{O}(10^5)$. We found that further reducing either the actuation frequency or the number of agents per simulation reduced the model's adaptability and therefore exhibit slightly lower performance.

Each LES is initialized for uniformly sampled $Re_\lambda \in \left\{65,\,76,\,88,\,103,\,120,\,140,\,163\right\}$ and a random velocity field synthesized from the target DNS spectrum. The residual-stress tensor $\tau^R$ is updated with equation~\ref{eq:ssm} and agents' actions every $\Delta t_{RL}$. The LES are interrupted at $T_{end} = 20 \tau_\mathrm{I}$ (between 750, if $Re_\lambda=65$, and 1600, if $Re_\lambda=163$, actions per agent) or if $\|\mathbf{u}\|_\infty > 10^3 u_\eta$, which signals numerical instability. The policy $\pi_{\mathtt w}$ is parameterized by a NN with 2 hidden layers of 64 units each, with $\tanh$ activations and skip connections. The NN is initialized as in~\cite{glorot2010} with small outer weights and bias shifted such that the initial policy is approximately $\pi_{\mathtt w^{(0)}}(\cdot|s) \approx \mathcal{N}(0.04, 10^{-4})$ and produces Smagorinsky coefficients with small perturbations around $C_s \approx 0.2$.
Gradients are computed with Monte Carlo estimates with sample size $B=512$ from a RM of size $N_{RM} = 10^6$. The parameters are updated with the Adam algorithm~\citep{kingma2014} with learning rate $\eta = 10^{-5}$. Each training run is advanced for $10^7$ policy gradient steps.
As discussed in the main text, because we use conventional RL update rules in a multi-agent setting, single parameter updates are imprecise.
We found that ReF-ER with hyper-parameters $C=1.5$ (Eq.~\ref{eq:refer}) and $D=0.05$ (Eq.~\ref{eq:refer beta}) to stabilize training.
We ran multiple training runs per reward function and whenever we vary the hyper-parameters, but we observe consistent training progress regardless of the initial random seed. The trained policies are evaluated by deterministically setting actions equal to the mean of the Gaussian $a(\mathbf{x}, t) = \mu_{\mathtt w}\left(s(\mathbf{x}, t)\right)$, rather than via sampling, and integrated in time for $100\tau_I$.

\textbf{Forced Isotropic Turbulence} \label{FHIT:TKE}
A turbulent flow is \emph{isotropic} when the averaged quantities of the flow are invariant under arbitrary translations and rotations. The flow statistics are independent of space and the mean velocity of the flow is zero. 
Forced, isotropic turbulence is governed by the incompressible Navier-Stokes equations,
\begin{align}
	\begin{cases}
		\frac{\partial \bm{u}}{\partial t} + \left( \bm{u} \cdot \nabla \right) \bm{u} &= - \nabla p + \nabla \cdot \left( 2 \nu S \right) + \bm {f} \\
		\nabla \cdot \bm{u} &= 0
	\end{cases}
\end{align}
where  $S = \frac{1}{2}(\nabla \bm{u} + \nabla \bm{u}^T)$ is the rate-of-strain tensor. The the turbulent kinetic energy (TKE, the second order statistics of the velocity field) is expressed as:
\begin{equation}
	e(\bm{x},t) \equiv \frac{1}{2}\bm{u}\cdot \bm{u}, \;\;  K(t) \equiv \frac{1}{2}\left < \bm{u}\cdot \bm{u}\right>,
\end{equation}
where the angle brackets $\left<\cdot\right> \equiv \frac{1}{\mathcal{V}}\int_\mathcal{D} \cdot$ denote an ensemble average over the domain $\mathcal{D}$ with volume $\mathcal{V}$.  
%\begin{equation*}
%	\frac{1}{2}\partial_t u_i^2 + u_i u_j \partial_j u_i = - u_i \partial_i p + \nu u_i \partial_j^2 u_i + u_i f_i
%\end{equation*}
%\begin{equation*}
%	\frac{\partial e}{\partial t}= - \nabla \cdot \left(\bm{u} p + \bm{u} e \right) + \nu \bm{u} \cdot \Delta \bm{u} +\bm{u} %\cdot \bm{f},
%\end{equation*}
%then we obtain the energy budget by integrating over $\mathcal{D}$. The surface terms on the boundary $\partial \mathcal{D}$ is zero for a flow field with \emph{periodic boundary conditions}:
For a flow with periodic boundary conditions the evolution of the kinetic energy is described as: 
\begin{align}
	\frac{dK}{dt} &=  - \nu \int_\mathcal{D}\|\nabla \bm{u}\|^2   + \int_\mathcal{D} \bm{u} \bm{f}  ~
	=  -2\nu \left< Z \right>\, +\, \left<\bm{u} \cdot \bm{f}\right>
\end{align}
where the energy dissipation due to viscosity, is  expressed in term of the norm of the vorticity $\bm{\omega} \equiv \nabla \times \bm{u}$ and the \emph{enstrophy}  $Z=\frac{1}{2}\bm{\omega}^2$.  This equation clarifies that  the vorticity of the flow field is responsible for energy dissipation that can only be conserved if there is a source of energy.

We investigate the behaviour of isotropic turbulence in a statistically stationary state by injecting energy through forcing.  In generic flow configurations the role of this forcing is taken up by the  large-scale structures and it is assumed that it does not influence smaller scale statistics, which are driven by viscous dissipation. The injected energy is transferred from large-scale motion to smaller scales due to the non-linearity of Navier-Stokes equations.
We implement a classic low-wavenumber (low-$k$) forcing term~\cite{ghosal1995} for isotropic turbulence that is proportional to the local fluid velocity as filtered from its large wave number components:
\begin{equation}
	\tilde{\bm{f}}(\bm{k},t) \equiv \alpha ~ G(\bm{k}, k_f) ~ \tilde{\bm{u}}(\bm{k},t) = \alpha ~ \tilde{\bm{u}}_{<}(\bm{k},t),
\end{equation}
where the tilde symbol denotes a three-dimensional Fourier transform, $G(\bm{k}, k_f)$ is a low-pass filter with cutoff wavelength $k_f$, $\alpha$ a constant, and $\tilde{\bm{u}}_{<}$ is the filtered velocity field. By applying Parseval's theorem, the rate-of-change of energy in the system due to the force is:
\begin{align}
	\left< \bm{f} \cdot \bm{u} \right> &= \frac{1}{2} \sum_{\bm{k}} \left( \tilde{\bm{f}}^*\cdot\tilde{\bm{u}} +  \tilde{\bm{f}}\cdot\tilde{\bm{u}}^* \right) = \alpha \sum_{\bm{k}} \tilde{\bm{u}}_<^2 = 2\alpha K_<.
\end{align}
Here, $K_<$ is the kinetic energy of the filtered field.
We set $\alpha = \epsilon/ 2K_<$ and $k_f = 4 \pi / L$, meaning that we simulate a time-constant rate of energy injection $\epsilon$ which forces only the seven lowest modes of the energy spectrum. The constant injection rate is counter-balanced by the viscous dissipation $\epsilon_{visc} = 2\nu \left<Z\right>$, the dissipation due to the numerical errors  $\epsilon_\mathrm{num}$, and, by a subgrid-scale (SGS) model of turbulence ($\epsilon_\mathrm{sgs}$, when it is employed - see Sec.~\ref{FHIT:Large Eddy Simulations}).  
When the statistics of the flow reach steady state, the time-averaged total rate of energy dissipation $\epsilon_\mathrm{tot} = \epsilon_{visc} + \epsilon_\mathrm{num} + \epsilon_\mathrm{sgs}$ is equal to the rate of energy injection $\epsilon$.

\textbf{The Characteristic Scales of Turbulence}  \label{FHIT:scales}
Turbulent flows are characterized by a large separation in temporal and spatial scales and long-term dynamics. %As mentioned in the previous paragraph, the turbulent properties of a flow arise from energy transfers from large scales (the physical boundaries of the flow) to smaller scales.
These scales can be estimated by means of dimensional analysis, and can be used to characterize turbulent flows. At the \emph{Kolmogorov scales} energy is dissipated into heat: $\eta = \left({\nu^3}/{\epsilon}\right)^{1/4}$, $\tau_\eta = \left({\nu}/{\epsilon}\right)^{1/2}$, $u_\eta = \left(\epsilon \nu \right)^{1/4}$.
These quantities are independent of large-scale effects including boundary conditions or external forcing. The \emph{integral scales} are the scales of the largest eddies of the flow:
$l_\mathrm{I} = \frac{3\pi}{4K} \int_0^\infty \frac{\tilde{E}(k)}{k} \mathrm{d}k$, $\tau_\mathrm{I} = \frac{l_\mathrm{I}}{\sqrt{2 K/3}}$.
The Taylor-Reynolds number is used to characterize flows with zero mean bulk velocity: $Re_\lambda = K \sqrt{ {20}/({3\nu \epsilon})}$.

Under the assumptions of isotropic flow we study the statistical properties of turbulence in Fourier space. We analyze quantities computed from simulations at statistically steady state and we omit the temporal dependencies. The energy spectrum is
$\tilde{E}(k) \equiv \frac{1}{2}\tilde{\bm{u}}^2(k)$.
Kolmogorov's theory of turbulence predicts the well-known $-\frac{5}{3}$ spectrum (i.e. $\tilde{E}(k) \propto \epsilon^{2/3} k^{-5/3}$) for the turbulent energy in the inertial range $ k_\mathrm{I} \ll k \ll k_\eta$.

\textbf{Direct Numerical Simulations (DNS)}
Data from DNS serve as reference for the SGS models and as targets for creating training rewards for the RL agents. The DNS are carried out on a uniform grid of size $512^3$ for a periodic cubic domain $(2\pi)^3$. The solver is based on finite differences, third-order upwind for advection and second-order centered differences for diffusion, and pressure projection~\cite{chorin1967}. Time stepping is performed with second-order explicit Runge-Kutta with variable integration step-size determined with a Courant–Friedrichs–Lewy (CFL) coefficient $CFL=0.1$.
We performed DNS for Taylor-Reynolds numbers in log increments between $Re_\lambda \in [60,\,205]$ (Fig.~\ref{fig:dns all}\textbf{a,b}). 

\begin{figure*}
	\includegraphics[width=\textwidth]{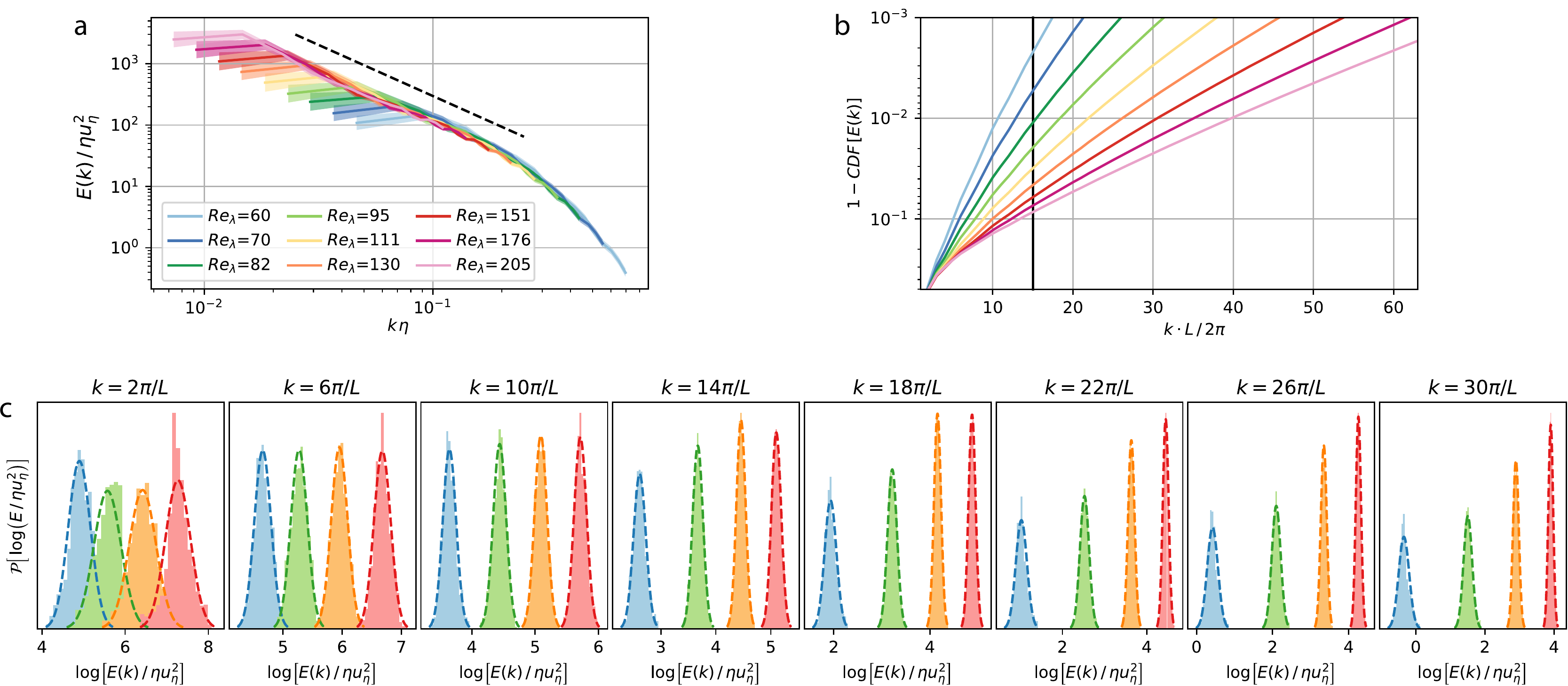}
	\caption{\textbf{Statistical properties of DNS simulations of forced isotropic turbulence.} 
		\textbf{a}, Time averaged energy spectra, and intervals of one standard deviation for log-increments of $Re_\lambda \in [60, 205]$ compared to Kolmogorov's spectrum $\propto k^{-5/3}$ (dashed line).
		\textbf{b}, Cumulative fraction of the TKE contained up to mode $k$. The black vertical line corresponds to the Nyquist frequency for the grid size ($N = 32^3$) used for all LES considered throughout this study.
		\textbf{c}, Histograms of single modes of the energy spectrum for $Re_\lambda = 65$ (blue),  88 (green), 110 (orange), and 163 (red) compared to a Gaussian fit.}
	\label{fig:dns all}
\end{figure*}
The initial velocity field is synthesized by generating a distribution of random Fourier coefficients matching a radial target spectrum $\tilde{E}(k)$~\cite{rogallo1984}:
$
\tilde{E}(k) = c_k ~ \epsilon^{2/3} k^{-5/3} ~ f_L(kL) ~ f_\eta(k\eta), \label{eq:pope spectrum}
$
where $f_l(kl_\mathrm{I})$ and $f_\eta(k\eta)$ determine the spectrum in the integral- and the dissipation-ranges respectively~\cite{pope2001}.
The choice of initial spectrum determines how quickly the simulation reaches statistical steady state, at which point $Re_\lambda$ fluctuates around a constant value. 
The time-averaged quantities (Fig.~\ref{fig:dns all}) are computed from 20 independent DNS with measurements taken every $\tau_\eta$. Each DNS lasts 20 $\tau_\mathrm{I}$ and the initial 10 $\tau_\mathrm{I}$ are not included in the measurements, which found to be ample time to avoid the initial transient.
Figure~\ref{fig:dns all}\textbf{c} shows that the distribution of energy content for each mode $\tilde{E}(k)$ is well approximated by a log-normal distribution such that $\log E^{Re_\lambda}_{DNS} \sim \mathcal{N}\left(\mu^{Re_\lambda}_{DNS} ,~ \Sigma^{Re_\lambda}_{DNS} \right)$, where $\mu^{Re_\lambda}_{DNS}$ is the empirical average of the log-energy spectrum for a given $Re_\lambda$ and $\Sigma^{Re_\lambda}_{DNS}$ is its covariance matrix.

\textbf{Large-Eddy Simulations (LES)} \label{FHIT:Large Eddy Simulations}
LES~\cite{leonard1974} resolve the  large scale dynamics of turbulence and model their interaction with the sub grid-scales (SGS).
The flow field $\bar{\bm{u}}$ on the grid is viewed as the result of filtering out the residual small-scales of a latent velocity field $\bm{u}$.
The filtered Navier-Stokes equation for the field $\bar{\bm u}$ reads:
\begin{equation}
	\frac{\partial \bar{\bm{u}}}{\partial t} + \left( \bar{\bm{u}} \cdot \nabla \right) \bar{\bm{u}} = - \nabla \bar{p} + \nabla \cdot \left( 2 \nu \bar{S}  - \tau^R\right) + \bar{\bm {f}} \label{eq:filtered navier stokes}
\end{equation}
here, the \emph{residual-stress tensor} $\tau^R$ encloses the interaction with the unresolved scales:
\begin{equation}
	\tau^R = \overline{\bm u \otimes \bm u} - \bar{\bm u} \otimes \bar{\bm u}. \label{eq:residual stress tensor}
\end{equation}
Closure equations are used to model the sub grid-scale motions represented by $\overline{\bm u \otimes \bm u}$.

\textbf{The Classic Smagorinsky Model (SSM)} The SSM~\citep{smagorinsky1963} is a linear eddy-viscosity model that relates the residual stress-tensor to the filtered rate of strain
\begin{align}
	\tau^R - \frac{1}{3}\textrm{tr} \left( \tau^R \right) &= -2\,\nu_t\,\bar{S}, \label{eq:smag} \\
	\nu_t &= \left( C_s \Delta \right)^2 \,\|\bar{S}\|, \label{eq:ssm}
\end{align}
where $\Delta$ is the grid size and $C_s$ is a constant.
This model has been shown  to perform reasonably well for isotropic turbulence and wall-bounded turbulence.
Eq.~\ref{eq:smag} models energy transfer from the filtered motions to the residual motions proportional to the turbulent eddy-viscosity $\nu_t$.
The main drawback of this model is that the constant $C_s$ has to be manually tuned. %the model introduces generally too much dissipation.

\textbf{The Dynamic Smagorinsky Model (DSM)} The DSM~\citep{germano1991} computes the parameter $C_s(\bm{x}, t)$ as a function of space and time. DSM's dynamic model is obtained by filtering equation~\ref{eq:filtered navier stokes} a second time with a so-called test filter of size $\widehat{\Delta} > \Delta$. The resolved-stress tensor $\mathcal{L}$ is defined by the Germano identity:
\begin{equation}
	\mathcal{L}_{\overline{\bm u}} = \reallywidehat{{\overline{\bm u} \otimes \overline{\bm u}}} - \widehat{{\overline{\bm u}}} \otimes \widehat{{\overline{\bm u}}} = T^R - \widehat{\tau^R}, \label{eq:germano identity}
\end{equation}
where $T^R = \reallywidehat{\overline{\bm u \otimes \bm u}} - 
\widehat{\overline{\bm u}} \otimes \widehat{\overline{\bm u}}$ is the residual-stress tensor for the test filter width $\widehat{\Delta}$, and $\widehat{\tau^R}$ is the test-filtered residual stress tensor for the grid size $\Delta$ (Eq.~\ref{eq:residual stress tensor}). If both residual stresses are approximated by a Smagorinsky model, the Germano identity becomes:
\begin{equation}
	\mathcal{L}_{\overline{\bm u}} \approx 2 \, C^2_s(\bm{x}, t)  \, \Delta^2  \, \left[ \reallywidehat{\|\bar{S}\| \bar{S}} - \frac{\widehat{\Delta}^2}{\Delta^2} \|\widehat{\bar{S}}\| \widehat{\bar{S}}\right]. \label{eq:approx germano}
\end{equation}
The dynamic Smagorinsky parameter (Eq.~\ref{eq:approx germano}) forms an over-determined system for $C^2_s(\bm{x}, t)$, whose least-squares solution is~\citep{lilly1992}:
\begin{equation}
	C^2_s(\bm{x}, t) = \frac{\langle\mathcal{L}_{\overline{\bm u}},\,\mathcal{M}\rangle_F}{2\Delta^2 \, \| \mathcal{M} \|^2}, \label{eq:dynamic smagorinsky parameter}
\end{equation}
where $\mathcal{M} = \reallywidehat{\|\bar{S}\| \bar{S}} -  (\widehat{\Delta}/ \Delta )^2 \, \| \widehat{\bar{S}}\| \widehat{\bar{S}}$, and $\langle\cdot\rangle_F$ is the Frobenius product. Because the dynamic coefficient may take negative values, which represents energy transfer from the unresolved to the resolved scales, $C^2_s$ is clipped to positive values for numerical stability.

The fraction of TKE contained in the unresolved scales increases with $Re_\lambda$ and decreases with the grid size (Fig.~\ref{fig:dns all}\textbf{b}). For all LES considered in this study we employ a grid of size $N=32^3$ and time-stepping coefficient $CFL=0.1$. For the higer $Re_\lambda$, the SGS model accounts for up to 10\% of the total TKE. We employ second-order centered discretization for the advection and the initial conditions for the velocity field are synthesized from the time-averaged DNS spectrum at the same $Re_\lambda$~\cite{rogallo1984}. When reporting results from SSM simulation, we imply the Smagorinsky constant $C_s$ resulting from line-search optimization. LES statistics are computed from simulations up to $t = 100\tau_\mathrm{I}$, disregarding the initial $10\tau_\mathrm{I}$ time units. For the DSM procedure we employ an uniform box test-filter of width $\widehat{\Delta} = 2 \Delta$.

\textbf{MARL models}
We defined two MARL SGS models by the reward function they optimize. Both reward functions have the form $r(\bm{x}^{(i)}, t) = r^{grid}(t) + r'(\bm{x}^{(i)}, t)$.
The base reward is a measure of the distance from the target DNS spectrum, derived from the regularized log-likelihood (Eq.~\ref{eq:ref loglikelihood}):
\begin{equation}
	r^{grid}(t) = \exp {\left[ -\sqrt{ - \langle \widetilde{LL}\rangle(t)} \right]}.
\end{equation}
This regularized distance is preferred because a reward directly proportional to the probability $\mathcal{P}(\tilde{E}(t) | E^{Re_\lambda}_{DNS})$ (Eq.~\ref{eq:ref probability}) quickly vanishes to zero for imperfect SGS models and therefore yields too flat an optimization landscape.
The average LES spectrum is computed with an exponential moving average with effective window $\Delta t_{RL}$:
\begin{equation}
	\langle \widetilde{LL}\rangle (t) = \langle \widetilde{LL}\rangle(t{-}\delta t) + \frac{\delta t}{\Delta t_{RL}} \left( \widetilde{LL}(\tilde{E}(t) | E_{DNS}) - \langle \widetilde{LL}\rangle(t{-}\delta t) \right)
\end{equation}
The reward $r^{G}$ adds a local term to reward actions that satisfy the Germano identity (Eq.~\ref{eq:germano identity}):
\begin{equation}
	r^{G}(\bm{x}, t) = r^{grid}(t) - \frac{1}{u_\eta^4}\| \mathcal{L}_{\overline{\bm u}}(\bm{x}, t) - T^R(\bm{x}, t) + \widehat{\tau^R(\bm{x}, t)}\|^2 \label{eq:reward germano}
\end{equation}
here the coefficient $u_\eta^4$ is introduced for non-dimensionalization.
The reward $r^{LL}$ further rewards matching the DNS spectra:
\begin{equation}
	r^{LL}(t) = r^{grid}(t) + \frac{\tau_\eta}{\Delta t_{RL}}\left[\langle \widetilde{LL}\rangle (t) - \langle \widetilde{LL}\rangle (t{-}\Delta t_{RL})\right] \label{eq:reward LL}
\end{equation}
This can be interpreted as a non-dimensional derivative of the log-likelihood over the RL step, or a measure of the contribution of each round of SGS model update to the instantaneous accuracy of the LES.
}

\showmatmethods{} % Display the Materials and Methods section

\subsection*{Data availability}
All the data analyzed in this paper was produced with open-source software described in the Code Availability Statement.
Reference data and the scripts used to produce the data figures, as well as instruction to launch the reinforcement learning training and evaluate trained policies, are available on a GitHub repository (\url{https://github.com/cselab/MARL_LES}).

\subsection*{Code availability}
Both direct numerical simulations and large-eddy simulations were performed with the flow solver \texttt{CubismUP 3D} (\url{https://github.com/cselab/CubismUP_3D}). The data-driven SGS models were trained with the reinforcement learning library \texttt{smarties} (\url{https://github.com/cselab/smarties}). The coupling between the two codes is also available through GitHub (\url{https://github.com/cselab/MARL_LES}). 

\subsection*{Correspondence and requests for materials} should be addressed to P.K. E-mail: petros@ethz.ch

\subsection*{Competing interests} The authors declare no conflict of interest.

\acknow{
We are very grateful to Dr. Hyun Ji (Jane) Bae (Harvard University) and Prof. Anthony Leonard (Caltech) for insightful feedback on the manuscript, and to Dr. Jacopo Canton and Martin Boden (ETH Zurich) for valuable discussions throughout the course of this work. We acknowledge support by the European Research Council Advanced Investigator Award 341117. Computational resources were provided by Swiss National Supercomputing Centre (CSCS) Project s929.
}

\showacknow{}

\subsection*{Author contributions}
G.N. and P.K. designed the research; G.N. and H.L. wrote the simulation software, G.N., H.L. and P.K. carried out the research; G.N. and P.K. wrote the paper.

% Bibliography
\bibliography{rl_hit}

\end{document}